\documentclass[prb,twocolumn,showpacs,superscriptaddress]{revtex4-1}

\usepackage{graphicx}% Include figure files
\def \uudd {\uparrow \uparrow \downarrow \downarrow}

\begin{document}
\title{Magnetic orders and spin-flop transitions in the cobalt doped
multiferroic $\rm Mn_{1-x}Co_{x}WO_4$}

\author{Feng Ye}
\affiliation{Quantum Condensed Matter Division, Oak Ridge National Laboratory,
Oak Ridge, Tennessee 37831, USA}
\affiliation{Department of Physics and Astronomy, University of Kentucky,
Lexington, Kentucky 40506, USA}
\author{Songxue Chi}
\affiliation{Quantum Condensed Matter Division, Oak Ridge National Laboratory,
Oak Ridge, Tennessee 37831, USA}
\author{Jaime~A.~Fernandez-Baca}
\affiliation{Quantum Condensed Matter Division, Oak Ridge National Laboratory,
Oak Ridge, Tennessee 37831, USA}
\affiliation{Department of Physics and Astronomy, University of Tennessee,
Knoxville, Tennessee 37996, USA}
\author{Huibo~Cao}
\affiliation{Quantum Condensed Matter Division, Oak Ridge National Laboratory,
Oak Ridge, Tennessee 37831, USA}
\author{K.-C. Liang}
\author{Yaqi Wang}
\author{Bernd Lorenz}
\affiliation{
Department of Physics and TCSUH, University of Houston, Houston, Texas 77204,
USA}
\author{C.~W.~Chu}
\affiliation{
Department of Physics and TCSUH, University of Houston, Houston, Texas 77204,
USA}
\affiliation{Lawrence Berkeley National Laboratory, 1 Cyclotron Road,
Berkeley, CA 94720, USA}
\date{\today}

\begin{abstract}
We present a comprehensive single-crystal neutron diffraction investigation of
the $\rm Mn_{1-x}Co_{x}WO_4$ with $0.02\leq x \leq0.30$. At lower
concentration $x \leq 0.05$, the system is quickly driven into the
multiferroic phase with spin structure forming an elliptical spiral order
similar to the parent compound. The reduction of electric polarization is
ascribed to the tilting of the spiral plane. For $x\sim 0.075$, the magnetic
structure undergoes a spin-flop transition that is characterized by a sudden
rotation of the spin helix envelope into the $ac$ plane. This spin structure
persists for concentration up to $x=0.15$, where additional competing magnetic
orders appear at low temperature.  For $0.17 \leq x \leq 0.30$, the system
experiences another spin-flop transition and recovers the low-$x$ spiral spin
configuration. A simple commensurate spin structure with $\vec{q}=(0.5,0,0)$
is found to coexist with the incommensurate spiral order. The complex
evolution of magnetic structure in Co doped $\rm MnWO_4$ contrasts sharply
with other transition metal ion doped $\rm Mn_{1-x}A_xWO_4$ (A=Zn, Mg, Fe)
where the chemical substitutions stabilize only one type of magnetic
structure. The rich phase diagram of $\rm Mn_{1-x}Co_{x}WO_4 $ results from
the interplay between magnetic frustration and spin anisotropy of the Co ions.
\end{abstract}

\pacs{75.30.Kz,75.25.-j,61.05.F-,75.58.+t}
%75.30.Kz, magnetic phase boundaries
%75.25.-j, magnetic structure determination
%61.05.F-, neutron diffraction
%75.58.+t, multiferroics

\maketitle

\section{Introduction}

The observation of spontaneous electric polarization and magnetic control of
ferroelectricity in perovskite manganite TbMnO$_3$\cite{kimura03} has
inspired much theoretical and experimental efforts searching for new
magnetoelectric multiferroic materials due to their great technological and
fundamental importance.\cite{khomskii06,tokura07,cheong07} New magnetic
multiferroics among transition metal oxides have since been discovered that
include rare-earth manganite derivative $R$MnO$_3$ and $R$Mn$_2$O$_5$ ($R$: Y, Gd,
Dy),\cite{goto04,kimura05,higashiyama04,hur04} geometrically-frustrated
triangular lattice Cu$T$O$_2$ ($T$: Fe,Cr),\cite{kimura06,kimura09} kagome
lattice antiferromagnet $\rm Ni_3V_2O_8$.\cite{lawes05}  In the conventional
(type-I) ferroelectric materials, the electric polar state arises either
from the covalent bonding between filled-shell oxygen atoms and the empty $d$-shell
nonmagnetic transition metal ion ({\it e.g.},~BaTiO$_3$),\cite{cohen92} or the
inversion symmetry breaking caused by the 6$s^2$ orbital (lone pair) that
moves away from the centrosymmetric position ({\it
e.g.},~BiMnO$_3$).\cite{seshadri01} Even if magnetic ions are present in
these materials, the spins order at much lower temperature than the electric
dipoles and the effect of magnetic transition on the dielectric constant is
weak. In contrast, the new family of magnetoelectric multiferroics (type-II)
exhibits an exceptionally strong sensitivity to an applied magnetic field that
causes reversal and sudden flops of the electric
polarization.\cite{kimura03,kimura05,hur04} Such a level of control indicates
the electric polarization is induced by the magnetic order, which typically
has an incommensurate noncollinear spiral configuration. The onset of
ferroelectricity correlates with the transition to the spiral spin order. The
intimate link between these two order parameters marks the prominent and
intriguing feature of the new class of multiferroics.

The mineral h\"{u}bnerite MnWO$_4$ (monoclinic, {\it P}2/{\it c}) is one of
the few multiferroics that is ideal for studying the complex spin orders caused by
magnetic frustrations and the interplay between magnetism and
ferroelectricity.\cite{taniguchi06,arkenbout06,heyer06} Without chemical
doping, the parent compound undergoes sequential magnetic transitions in zero
magnetic field.\cite{lautenschlager93,ehrenberg97} The system first enters
a collinear spin structure (AF3) around 13.5~K with sinusoidal modulation of the
magnetic moment, the corresponding incommensurate (ICM) wave vector appears at
$\vec{q}_3=(0.214,0.5,-0.457)$ and the moment of the Mn ions are confined in
the $ac$ plane at an angle of 35$^{\circ}$ towards the $a$ axis. At 12.6~K, an
ICM elliptical spiral spin order (AF2) sets in, accompanied by the spontaneous
electric polarization along the crystallographic $b$ axis. The magnetic order
at AF2 phase has the same magnetic wave vector $\vec{q}_2=\vec{q}_3$ while the
helix spin structure is characterized by the moment tilting out of the $ac$
plane towards the $b$ axis. As the system is further cooled below 7~K,
the ICM magnetic order is replaced by a commensurate (CM) magnetic order (AF1)
that also suppresses the electric polarization. The close proximity of three
different magnetic phases and the metamagnetic transitions prove the
existence of significant magnetic frustration in the system, as revealed by
inelastic neutron scattering measurements.\cite{ehrenberg99,ye11}
Consistent with theoretical studies,\cite{tian09,matityahu12},
the investigation of spin wave excitations in the low-$T$ $\uudd$ spin order
indicates that the collinear spin state results from the balance of long range
magnetic interactions between Mn$^{2+}$ ions up to 11$^{\rm th}$
neighbors.\cite{ye11} Higher-order magnetic exchange interactions have
strengths comparable to the nearest neighbor exchange coupling along the
zigzag spin chain in the $c$ axis and between the zigzag chains in the $a$
axis direction.  However, the magnetic interactions between the chains in the $b$
axis are much weaker. This finding suggests that different magnetic structures
are close in energy and compete for the magnetic ground state. Like other
multiferroic materials, the magnetic and ferroelectric phases in this system
are expected to be tunable with small perturbations of magnetic
field,\cite{taniguchi06,higashiyama04,hur04,seki08}
pressure,\cite{cruz07,chaudhury08,cruz08,chaudhury07} and even chemical
substitution with various transition metal ions. MnWO$_4$ is known to form
stable compounds when Mn is replaced by Fe,\cite{kleykamp80,matres03} Co, Ni,
Cu,\cite{weitzel70} or Zn.\cite{takagi81} It was reported that the chemical
doping with magnetic Fe ions stabilize the low-$T$ collinear and CM spin
order.\cite{ye08,chaudhury09,liang11} Both neutron diffraction and electric
polarization measurements show that the low-$T$ magnetic ground state is
completely converted into the collinear spin order (AF1 phase) with 5\% of Fe
substitution, which is largely due to the increased Fe single ion
anisotropy.\cite{ye12} In contrast, substitution with nonmagnetic ions (Zn or
Mg) has been shown to alter the magnetic ground state differently; replacement
of only a few percent of nonmagnetic ions seems to be
sufficient to suppress the commensurate AF1 phase and stabilize the spiral
spin order.\cite{meddar09,chaudhury11}  A similar effect was also observed in
the Co-substituted compound $\rm Mn_{1-x}Co_{x}WO_4$ by neutron diffraction
and bulk magnetic measurements on polycrystalline samples. Song {\it et
al.}\cite{song09} found that the CM collinear state is replaced by the ICM
spiral phase with 5\% Co concentration; a spin-flop transition occurs with increasing Co doping with the corresponding spiral
plane tilting away from the $b$ axis.  The modification of the magnetic
structure is expected to switch the electric polarization direction from the $b$
to the $a$ axis as it was later confirmed in a single crystal of $\rm
Mn_{0.9}Co_{0.1}WO_4$.\cite{song10,olabarria12} However, it was discovered
that with further increasing of Co doping ($\sim$~15\%), the only measurable
polarization was along the $b$ axis.\cite{chaudhury10} This result indicates
that a new type of spin structure responsible for the $a$ axis polarization is
bound to a narrow range of Co substitution and the actual $x-T$ phase diagram
of the $\rm Mn_{1-x}Co_xWO_4$ is more complex than the one reported from
powder diffraction studies.\cite{song09}
\begin{figure}[ht!]
\includegraphics[width=3.1in]{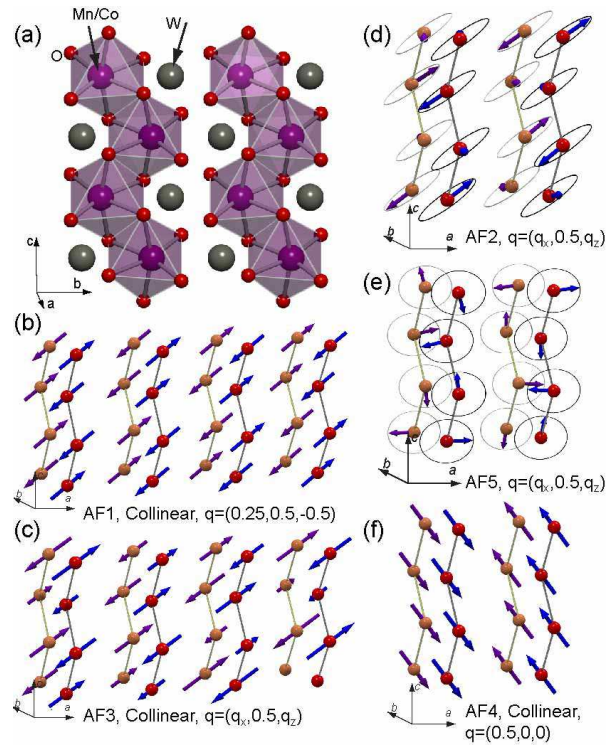}
\caption{(Color online) (a) The crystal structure of the $\rm
Mn_{1-x}Co_xWO_4$. The spin structures of (b) the collinear, commensurate (CM)
AF1 phase, spin moments are in the $ac$ plane
with an angle of $\sim 35^{\circ}$ to the $a$ axis, (c) the collinear,
incommensurate (ICM) AF3 phase, magnetic moments have the same
direction as AF1 but with modulated amplitude, (d) the noncollinear, ICM AF2
phase, one axis of the spiral ellipse lies in the $ac$ plane, the
other along the $b$ axis. (e) the noncollinear ICM AF5 phase, the
envelope of the spin helix lies in the $ac$ plane, (f) the collinear, CM AF4
phase, spin moments are in the $ac$ plane with a angle of
$-50^{\circ}$ to the $a$ axis. Per rotation convention for the right-handed
coordinates, the positive rotation angle with respect to the $a$ axis is
associated with the counterclockwise rotation when viewing the system along
the negative direction of the axis of rotation.
}
\label{fig:structure}
\end{figure}

\begin{figure}[ht!]
\includegraphics[width=3.4in]{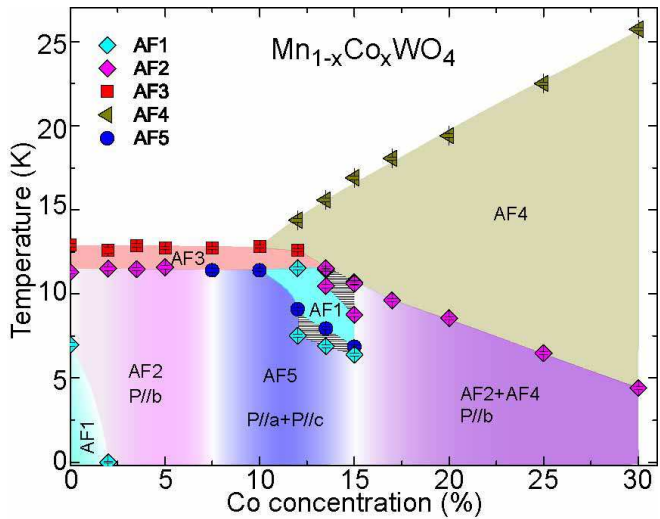}
\caption{(Color online) The phase diagram of $\rm Mn_{1-x}Co_{x}WO_4$ as
a function of temperature and Co concentration $x$. Due to the strong
hysteresis of the magnetic order, transitions between different phases are
identified with warming protocol. Labels of various magnetic phases correspond
to the spin structures displayed in Fig.~1.
}
\label{fig:PD}
\end{figure}

To elucidate the nature of the magnetic ground state upon Co doping, we
performed a comprehensive neutron single-crystal diffraction study of
the doped $\rm Mn_{1-x}Co_{x}WO_4$. We observed a systematic evolution
of the magnetic structure with increasing Co doping. The spin configurations
for different phases are depicted in Fig.~\ref{fig:structure}. For lower
concentration ($x \leq 0.05$), the system exhibits a spiral spin
configuration similar to the undoped MnWO$_4$, but with decreased angle
between the normal axis of the spiral plane and the $c$ axis. At $x\approx 0.075$, the
system undergoes a magnetic transition to a phase in which the spin helix flops
into the $ac$ plane.  The $ac$ spiral spin structure and the associated
polarization $P_a$ and $P_c$ survive in the range of $0.075 \leq x \leq
0.15$. For higher Co concentration, the system experiences a second spin-flop
transition such that the spin order switches back to low-$x$ spiral
structure coexisting with a collinear CM magnetic structure as observed in pure
CoWO$_4$.\cite{forsyth94} The complete phase diagram shown in Fig.~\ref{fig:PD} of
Co-substituted MnWO$_4$ is obtained based on the neutron diffraction and
bulk property measurements. The chemical substitution of the magnetic Co ion
provides an unique method to fine tune the magnetic property that is capable
of achieving magnetoelectric control with multiple value states.

\begin{figure}[ht!]
\includegraphics[width=3.3in]{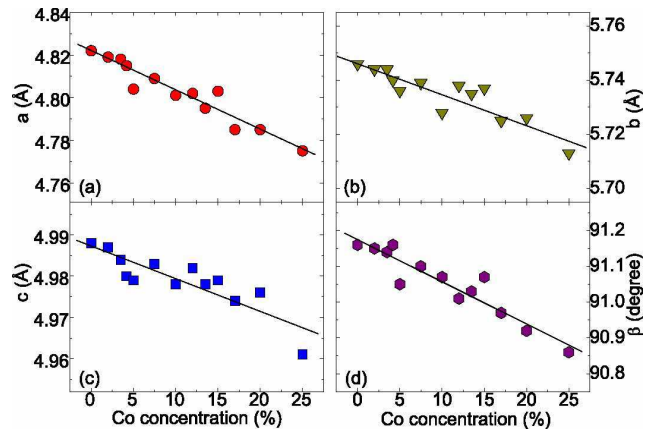}
\caption{(Color online) Doping dependence of the lattice parameter $a$,$b$,$c$
and the angle $\beta$ of $\rm Mn_{1-x}Co_{x}WO_4$ at room temperature
characterized by x-ray powder diffraction. The crystal structure of all
samples are refined with the monoclinic {\it P}2/{\it c} space group. Solid
lines are guides to the eye.
}
\label{fig:xray}
\end{figure}

Single crystals of $\rm Mn_{1-x}Co_xWO_4$ with thirteen different compositions
($ 0.02\leq x \leq 0.3$) have been grown in a floating zone optical furnace.
We use powder x-ray diffraction to check the phase purity of the
polycrystalline feed rod before the crystal growth. No impurity phase could be
detected within the resolution of the spectra. The chemical composition and
the Co content of the single crystals were verified by energy-dispersive x-ray
(EDX) measurements testing up to 15 different spots of a single crystal. As
further confirmed by neutron diffraction measurement, the refined Co content of all
samples was close to the nominal composition. In the following sections, we
will use the nominal composition to distinguish between different substitution
levels.  Single crystal neutron diffraction experiments were carried out at
the High Flux Isotope Reactor of the Oak Ridge National Laboratory. We used
the HB1A, HB1, and HB3 triple axis spectrometers to study the doping and
temperature evolution of the magnetic diffraction pattern. We chose an
incident neutron beam with wavelength of 2.366~${\rm \AA}$ and pyrolytic
graphite (PG) crystals as monochromator and analyzer. The crystals were
aligned in several scattering planes to probe different magnetic
reflections. For nuclear and spin structure determination, we used the HB3A
four-circle diffractometer to collect both the nuclear and
magnetic reflections with neutrons of wavelength 1.536 $\rm \AA$ at selected
temperatures. The crystal and magnetic structure refinement were performed
using the {\sc fullprof suite} package.\cite{fullprof} Magnetic representation
analysis is performed to choose appropriate basis vectors to describe the
various spin structures.  The sample temperature was regulated either using
a closed cycle refrigerator (CCR) or liquid Helium cryostat.  All
samples under study belong to the monoclinic {\it P}2/{\it c} (No.~13) space
group.  With increasing Co concentration $x$, the lattice parameters of
$a,b,c,$ and the angle of $\beta$ systematically decrease over the range
studied, as determined by the x-ray powder diffraction measurement shown in
Fig.~\ref{fig:xray}.

\section{magnetic order at low concentration ($0.02\leq x \leq 0.05$) }
\label{sec:lowX}

\begin{figure}[ht!]
\includegraphics[width=3.0in]{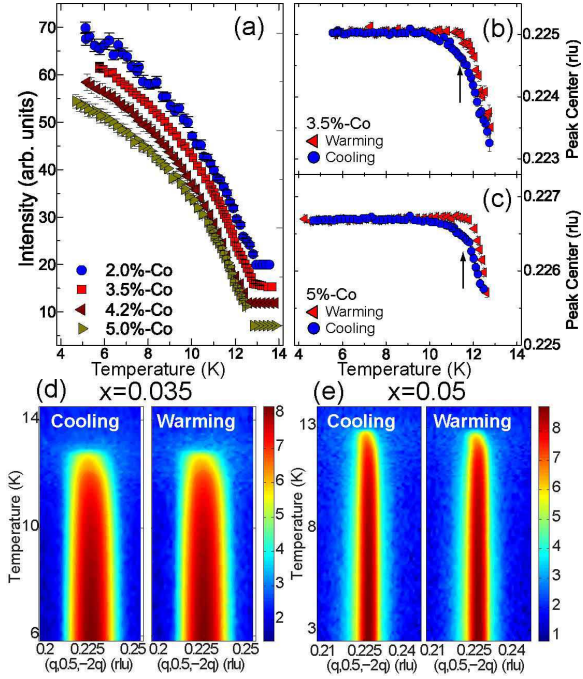}
\caption{(Color online) (a) The $T$ dependence of the ICM magnetic order
parameters for $\rm Mn_{1-x}Co_xWO_4$ with $x=0.02,0.035,0.042$, and $0.05$.
The order parameter data are shifted vertically for clarity. (b) and (c)
display the $T$ dependence of the ICM peak centers upon cooling and warming
for the $x=0.035$ and $x=0.05$ samples, respectively. Arrows indicate the
transition between the high-$T$ collinear and the low-$T$ noncollinear phases.
Contour plot of wave vector scans as a function of temperature for (d)
$x=0.035$ and (e) $x=0.05$ upon cooling and warming.
}
\label{fig:OP.lowX}
\end{figure}

We start with the neutron diffraction results of the lower Co concentration.
The samples were aligned in a horizontal scattering plane defined by two
orthogonal vectors of [1,0,-2] and [0,1,0]. Although the magnetic propagation
wave vectors in the pure MnWO$_4$ is determined to be
$\vec{q}=(0.214,0.5,-0.457)$ that deviates slightly from the scattering plane
used in the measurement, the coarse vertical resolution of the triple axis
spectrometer is sufficient to capture the magnetic Bragg reflections and track
their temperature and doping evolution. In the rest of the paper, we will use
the wave vector $(q,0.5,-2q)$, which is the projected value in the horizontal
scattering plane, to denote the ICM magnetic Bragg peak unless 
specified otherwise. Figure \ref{fig:OP.lowX} summarizes the thermal evolution of the
ICM magnetic order at lower cobalt concentration with $x=0.02, 0.035, 0.042$,
and 0.05.  The integrated intensities of magnetic scattering for these samples
grow monotonically below 13~K without any sign of anomaly at lower
temperatures.  Such behavior contrasts with the undoped MnWO$_4$ where the
multiferroic AF2 phase is sandwiched between the high-$T$ AF3 and the low-$T$
AF1 phases, resulting in an abrupt suppression of the ICM magnetic scattering.
Like the nonmagnetic Zn and Mg doping,\cite{meddar09,chaudhury11} the
multiferroic state associated with the noncollinear ICM phase is stabilized
with Co substitution and becomes the ground state. The
$T$ dependence of the integrated intensities does not separate the phase
boundary between the high-$T$ AF3 and the multiferroic AF2 phases
[see Fig.~\ref{fig:OP.lowX}(a)].  However, there are clear changes of the peak
center across the phase transition that label the phase boundary between the
collinear and noncollinear phases. In addition, the peak position is
independent of the temperature in the AF2 phase, but changes continuously in
the AF3 phase, as demonstrated in Figs.~\ref{fig:OP.lowX}(b) and
\ref{fig:OP.lowX}(c) for the $x=0.035$ and 0.05 samples.

\begin{table}[ht!]
\caption{Refined parameters of the magnetic structure for the
$x=0.02,0.035,0.042,$ and $0.05$ samples in the noncollinear AF2 phase. $m_b$
and $m_{\perp b}$ denote the projected moments along the two principle axes of
the spin ellipse along and perpendicular to the $b$ axis. $\theta$ is the
angle between $m_{\perp b}$ and the $a$ axis, or the angle between the normal
vector $\vec{n}$ of the spiral plane and the $c$ axis. The eccentricity of the
spin helix is defined as $\epsilon=(1-m^2_{b}/m^2_{\perp b})^{1/2}$ to
describe the deviation from a circular envelope. Notice the $x=0.075$ sample
undergoes rotation transition of the spiral plane on cooling and locks into
the AF2 phase.  $\rm R_{F^2}$ is the $R$ factor calculated by $R_{F^2} = 100
\sum_{n}(|G^2_{\rm obs}-G^2_{\rm calc}|)/\sum_{n}G^2_{\rm obs}$, where $G^2$
is the square of the structure factor for $n$ observed reflections.
}
\begin{ruledtabular}
\begin{tabular}{llccccc}
 nominal $x$ &  $m_b (\mu_B)$  & $m_{\perp b} (\mu_B)$  & $\epsilon$ &
 $\theta$ & ${\rm R_{F^2} (\%)}$&	\\
    \hline
  0.02  	& 3.86(5) & 4.44(6)&	0.49(4) 	& 28.4(9)	& 4.22 &   \\
  0.035 	& 3.91(5) & 4.42(6)&	0.47(4) 	& 19(2)		& 5.83 &   \\
  0.042 	& 3.86(6) & 4.52(9)&	0.52(5) 	& 20(1)		& 5.71 &   \\
  0.05  	& 3.82(5) & 4.33(6)&	0.47(4) 	& 15(1)		& 4.90 &   \\
  0.075 	& 3.47(5) & 4.62(6)&	0.66(3) 	& 6(1) 		& 7.56 &   \\
  \end{tabular}
\end{ruledtabular}
\label{tab:refine.lowX}
\end{table}

\begin{figure}[ht!]
\includegraphics[width=3.3in]{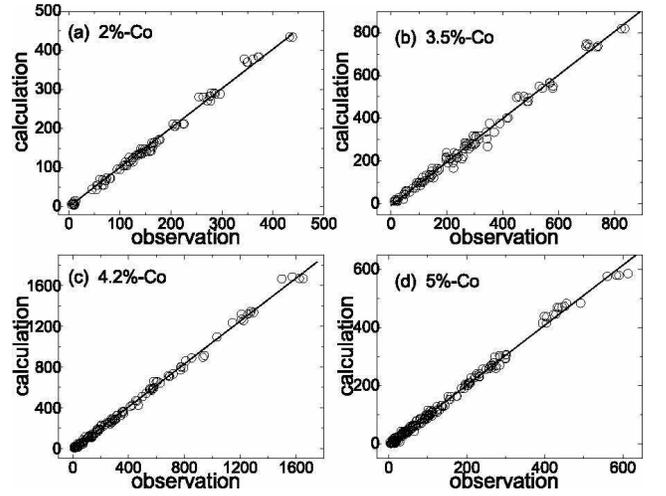}
\caption{Agreement of the magnetic structure refinements at (a) $x=0.02$,
(b) $x=0.035$, (c) $x=0.042$, and (d) $x=0.05$. The calculated structure
factors $F^{2}_{\rm calc}$ are plotted against the experimentally observed
$F^{2}_{\rm obs}$. The corresponding spin structures are shown in
Fig.~\ref{fig:structure}(d) and the inset of Fig.~\ref{fig:angle}.
}
\label{fig:refinement.lowX}
\end{figure}

The magnetic configurations of the AF2 phase for $\rm Mn_{1-x}Co_{x}WO_4$
at low $x$ can be determined using single crystal magnetic diffraction
experiments. As summarized in Table~\ref{tab:refine.lowX}, the refinement
parameters indicate the spin structure is well described by the elliptical
spiral as in the pure MnWO$_4$ and the doped $\rm
Mn_{1-x}Zn_{x}WO_4$.\cite{lautenschlager93,chaudhury11}
Figure~\ref{fig:refinement.lowX} shows the agreement between model
calculation and experimental observations for all samples. Contrary to the Zn
substitution, where the angle $\theta$ between the normal vector $\vec{n}$ of
the spiral plane and the crystallographic $c$ axis remains constant for Zn
concentration up to $x=0.40$, the value of $\theta$ shows a gradual decrease
with increasing Co concentration [see Fig.~{\ref{fig:angle}(a)].  The refinements
also indicate a considerable deviation of the ellipse from a
perfect circular envelope as quantified in the Table~{\ref{tab:refine.lowX}.
To correlate the spin structure and the ferroelectric properties,
Fig.~\ref{fig:angle}(b) displays the $T$ dependence of electric polarization
$\vec{P}$ at different $x$. Clearly, a smooth decrease of the saturated
polarization along the $b$ axis is observed, a trend similar to the doping
dependence of the angle $\theta$. Within the microscopic picture of spin
current or inverse Dzyaloshinskii-Moriya (DM)
model,\cite{katsura05,mostovoy06,sergienko06} the expected electric
polarization can be expressed as
\begin{equation}
\vec{P} = A \vec{e}_{ij} \times (\vec{S}_i \times \vec{S}_j),
\label{eqn:P}
\end{equation}
where $\vec{e}_{ij}$ is the unit vector connecting the neighboring spins
$\vec{S}_i$ and $\vec{S}_{j}$, and $A$ is a constant related to the
spin-orbit coupling and spin exchange
interaction.\cite{katsura05,mostovoy06,sergienko06} The inset of
Fig.~\ref{fig:angle} describes a general spin helix configuration; the normal
vector $\vec{n}$ of the spiral plane has angle $\theta$ with respect to the
$c$ axis and its projection in the $ab$ plane has angle $\phi$ towards the $a$
axis.  Using spherical coordinates, we can express $\vec{n}=\sin\theta\cos\phi
\vec{e}_x+\sin\theta\sin\phi \vec{e}_y+\cos\theta \vec{e}_z$, where
$\vec{e_i}$ are the unit vectors along the $i^{\rm th}$ Cartesian coordinates
($i=x,y,z$). Since the angle between the crystallographic $a$ and $c$ axes is
close to 90$^{\circ}$, we will consider $\vec{e}_x \parallel \vec{a}$,
$\vec{e}_y \parallel \vec{b}$, and $\vec{e}_z \parallel \vec{c}$ for the sake
of simplicity.  The magnetic moments of two spins in the chemical unit cell
for the helical state can be expressed as
\begin{equation}
m(\vec{R}_i,\alpha)=m_{\parallel}\cos(2\pi \vec{q}\vec{R}_i+\Phi_{\alpha})
+ m_{\perp}\sin(2\pi \vec{q}\vec{R}_i+\Phi_{\alpha}),
\end{equation}
where $\vec{R}_{i,\alpha}$ is the position vector of the Mn site
$\alpha(=1,2)$ in the unit cell $i$, $\vec{q}$ is the magnetic propagation
wave vector of the spiral structure, $\Phi_2=\Phi_1+\pi(q_z+1)$, and
$m_\parallel$ and $m_\perp$ are the long and short half axes of the magnetic
ellipse, respectively.  If we consider the dominant magnetic exchange coupling
is along the $c$ axis zigzag chain, the electric polarization can be
quantitatively written as
\begin{equation}
\vec{P}_{1} = C_1 m_{\parallel} m_{\perp}\sin\pi q_z \sin\theta (-\sin\phi \vec{e}_x + \cos\phi \vec{e}_y).
\label{eqn:angle}
\end{equation}
If, however, there is additional non-negligible exchange interactions between
the chains along the $a$ axis, the extra contribution to the polarization is
\begin{equation}
\vec{P}_{2} =C_2 m_{\parallel} m_{\perp}\sin2\pi q_x (-\cos\theta \vec{e}_y + \sin\theta \sin\phi
\vec{e}_z),
\end{equation}
where $C_1$, $C_2$ are constants independent of $\theta$, $\phi$ and only
related to the spin-orbit coupling and magnetic interaction, 
$q_x$, $q_z$ are components of the ICM magnetic wave vector.  The spin structure
refinement in the AF2 phase reveals $\phi=180$; this leads to a simplified form
of $\vec{P}_1=-C_1 m_{\parallel} m_{\perp} \sin\pi q_z\sin\theta\vec{e}_y$ and
$\vec{P}_2= -C_2 m_{\parallel} m_{\perp} \sin2\pi q_x\cos\theta\vec{e}_y$,
both contributing to the $b$ axis polarization. If there are
no significant deviations of the ellipticity and the moment size (Table~I), the
magnitude of $\vec{P}$ depends solely on the angle $\theta$.  With both
$\vec{P}_1$ and $\vec{P}_2$ taken into account, the total electric
polarization becomes $P_b \sim -m_{\parallel} m_{\perp}  (C_1\sin\theta + C_2
\cos\theta)$.  It is evident that the polarization will change from
$\vec{P}_1$ to $\vec{P}_2$ upon variation of $\theta$ from 90$^{\circ}$ to
0$^{\circ}$. In addition, we note that $|C_2|<|C_1|$ due to a shorter exchange
path along the zigzag chain direction ($3.29~\AA$ along the $c$ axis versus
$4.83~\AA$ along the $a$ direction), the expression can be further simplified
as $P_b \sim -C_1 m_{\parallel} m_{\perp} \sin \theta$. One can then estimate
that the saturated polarization $P_b(x=0.05)$ decreases to
$\sin15^\circ/\sin35^\circ\sim0.45 P_b(x=0)$ due to the rotation of the spiral
plane, which is close to the measured value. The close correlation between the
rotation of the spin helix plane and the decreasing ferroelectric polarization
emphasizes that the doped MnWO$_4$ is indeed a prototypical multiferroic
material with inverse DM mechanism being the origin of the spontaneous
electric polarization.

\begin{figure}[ht!]
\includegraphics[width=3.3in]{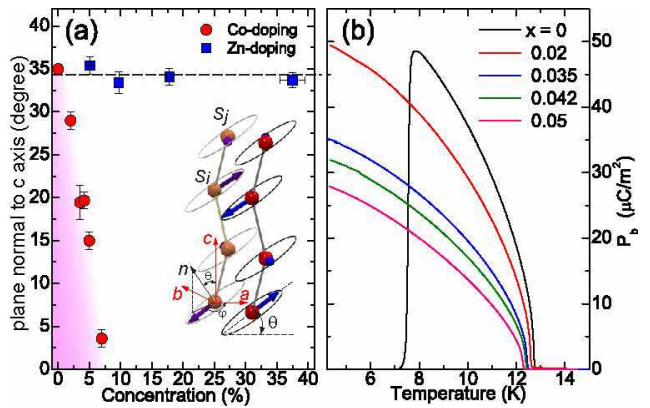}
\caption{(Color online) (a) The doping dependence of the angle $\theta$
between spiral plane normal and the $c$ axis for the Co doped MnWO$_4$. For
comparison, the results of the Zn-doped samples are also shown. Inset shows
the schematics of the elliptical spiral spin configuration in the AF2 phase.
The normal vector $\vec{n}$ of the spiral plane has a angle of $\theta$ with
respect to the $c$ axis, the projection of $\vec{n}$ on the $ab$ plane has a
angle of $\phi$ to the $a$ axis. (b) The $T$ dependence of electric
polarization $\vec{P_b}$ for $x=0,0.02,0.035,0.042$ and $0.05$ Co-doping.\cite{liang12} }
\label{fig:angle}
\end{figure}

\section{magnetic order at intermediate concentration ($0.075\leq x\leq 0.15$) }
\label{sec:middleX}

As the Co concentration increases to $x=0.075$, ferroelectric polarization
measurement does not detect any significant $b$ axis component within the
resolution of the experiment.\cite{liang12} However, a large polarization was
found along the $a$ axis accompanied by a small component along $c$, which
implies that the system undergoes major change in spin structure that causes
the reorientation of the polarization.  To correlate the polarization result,
we studied the $T$ dependence of the wave vector scans upon cooling and warming
as displayed in Figs.~\ref{fig:ICM.7p5Co}(a)-(b). One can clearly observe a
strong hysteresis in the magnetic scattering during thermal cycling. Upon
cooling, the peak center shifts from $\vec{q}_1 \approx
(0.234,0.5,-0.468)$ that is associated with the high-$T$ collinear AF3 phase
and moves into a plateau with another ICM wave vector $\vec{q}_2 \approx
(0.232,0.5,-0.464)$ for 7~K$<T<$11~K [Figs.~\ref{fig:ICM.7p5Co}(a) and
\ref{fig:ICM.7p5Co}(d)]. As the sample is further cooled to lower temperature,
the magnetic peak resumes shifting and finally locks into the low-$T$ ICM
phase with $\vec{q}_3=(0.229,0.5,-0.458)$.  On warming, the same low-$T$
magnetic order remains at the wave vector $\vec{q}_3$ till $T\sim$ 10~K and
suddenly is converted to the high-$T$ collinear AF3 phase.

\begin{figure}
\includegraphics[width=3.0in]{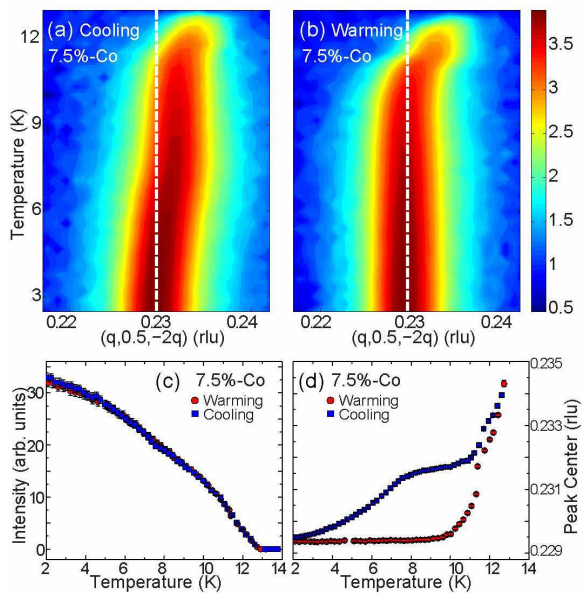}
\caption{(Color online) The $T$ dependence of the wave-vector scans of the
ICM magnetic peak for $x=0.075$ sample upon (a) cooling and (b) warming. The
dashed lines in (a)-(b) are guides the eyes to illustrate the locked in peak
position at the lowest temperature probed. (c)and (d) display the
$T$ dependence of the integrated intensity and the peak center of the ICM
scattering.
}
\label{fig:ICM.7p5Co}
\end{figure}

To understand the complex evolution of the magnetic order, the same sample
was placed on the four-circle diffractometer to investigate the
spin structure. We chose the cooling protocol and collected data at two
characteristic temperatures $T=9$~K and 4.5~K.
Figures.~\ref{fig:structure.7p5Co}(a) and \ref{fig:structure.7p5Co}(b) display
the refined spin configurations. At $T=9$~K, the magnetic spins form an $ac$
spiral structure (termed as AF5 phase) with two principle axes of the ellipse
lying along the crystallographic $a$ and $c$ directions. The normal vector
$\vec{n}$ of the new spiral plane is characterized by $\theta=90^{\circ}$ and
$\phi=90^{\circ}$. According to Eqn.~(\ref{eqn:P}), such a magnetic structure
will cause the rotation of $P_b$ into other directions. If we only consider
that the dominant magnetic interactions are along the spin chain direction
($c$ axis), the cross product of $\vec{n}$ and $\vec{e}_{ij}$ would only
induce the $a$ axis electric polarization in a form of 
\begin{equation}
P_a =-C_1 m_a m_c \sin \pi q_z \vec{e}_x,
\label{eqn:a_P}
\end{equation}
where $m_a$ and $m_c$ are the projected moments of the spin helix along $a$
and $c$.  This prediction only agrees partially with the experimental
observation where both $P_a$ and $P_c$ develop below 10~K.\cite{liang12} To
explain the presence of $P_c$, we have to consider the interchain magnetic
interactions along the $a$ axis. Recent spin wave measurements in MnWO$_4$
have revealed that the interchain magnetic interactions are of the same order
as the intrachain exchange coupling in its collinear spin state.\cite{ye11}
Although the results obtained are for the collinear spin structure, it is
expected they will have similar strength when the system enters the
noncollinear AF2 phase, based on what has been found in the pure CuFeO$_2$ and
the multiferroic $\rm CuFe_{1-x}Ga_{x}O_2$.\cite{ye07,haraldsen10} Taking the
interchain magnetic interactions into account, the contribution to the
electric polarization has the form of 
\begin{equation}
P_c =C_2 m_a m_c \sin2\pi q_x\vec{e}_z. 
\label{eqn:c_P}
\end{equation}
This prediction of small polarization along $c$ is in accordance with the bulk
polarization measurement. We note that although the spins of the AF5 phase
remain in the $ac$ plane, the expressions of Eqns.~\ref{eqn:a_P} and
\ref{eqn:c_P} indicate a $T$-dependent eccentricity that can further modify the
magnitude of the electric polarization. 

\begin{figure} \includegraphics[width=3.3in]{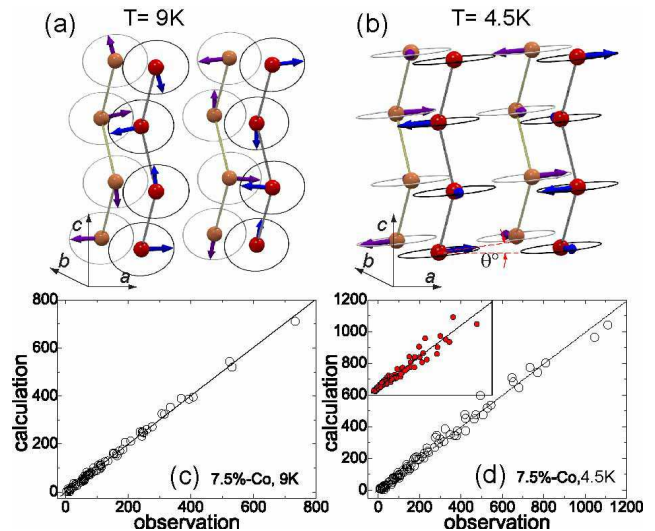}
\caption{(Color online) The refined spin structure for $x=0.075$ sample at (a)
9~K and (b) 4.5 K. The corresponding agreement of the refinements are
plotted on (c) and (d). Inset of panel (d) shows the agreement plot at
$T=4.5$~K using the same elliptical spiral at 9~K. The scattered
data points indicate a poorer description of the spin configuration.
}
\label{fig:structure.7p5Co}
\end{figure}

With the sample cooled below 8~K, it was found that both $P_a$ and $P_c$ are
suppressed. Such behavior suggests a gradual change of the spin structure that
might be due to either the distortion of the elliptical envelope or the
rotation of the spiral plane. Since both polarizations are proportional to
$m_a  m_c$ for an $ac$ spiral structure, the increased eccentricity of the
spiral ellipse could certainly yield a reduced bulk polarization.  On the
other hand, the rotation of the spiral plane away from the $ac$ plane can also
lead to the suppression of polarization along both directions. The spin
structure determination with data collected at 4.5~K provides a definitive
answer to separate those two possibilities. As shown in
Fig.~\ref{fig:structure.7p5Co}(b), the system at base temperature shows a
completely different spin structure from the one at 9~K.  It possesses the
configuration similar to the low Co concentration with the short axis of the
spin ellipse along the $b$ axis and the long axis residing in the $ac$ plane.
The canted angle $\theta$ between the spiral plane normal vector $\vec{n}$ and
the $c$ axis is $\sim~6^\circ$.  The result reveals that the $x=0.075$
sample is located near the boundary between distinct spin configurations of
the low-$x$ spiral and the $ac$ spiral structure.

To understand the spin flop transition near $x_c=0.075$, we recall that the
spiral spin order allows the coupling linear in gradient of the magnetic order
parameter (also known as Lifshitz invariant) with broken inversion symmetry and
induces a uniform electric polarization according to the Ginzburg-Landau
approach.\cite{mostovoy06} The energy gain originating from the nonlinear
coupling between $\vec{P}$ and $\vec{M}$ ($\sim \gamma
{\vec{P}\cdot[\vec{M}(\nabla \cdot \vec{M})-(\vec{M}\cdot\nabla)\vec{M}]}$) and
the corresponding electric polarization is proportional to the angle
$\theta$. With increasing Co concentration, both the electric polarization
$\vec{P}$ and the energy gain decrease because of the decreasing angle. It is
not surprising that exists a critical concentration $x_c$, around which
the system can no longer gain energy to maintain the magnetic structure. Such
instability will then bring about a spin flop transition to the observed $ac$
spiral which can reduce the free energy further because of the sizable
polarization.  Compared to other multiferroics, the continuous change in the
magnetic wave vector from the transition between different spin configurations of
the Co doped MnWO$_4$ appears rather rare. The gradual transition highlights
that the high-$T$ spiral and the low-$T$ $ac$ spiral spin structures are
nearly degenerate in energy such that a small change in temperature would
drive the system between competing magnetic states characterized by a
continuous rotation of the spiral plane.

\begin{figure}
\includegraphics[width=3.0in]{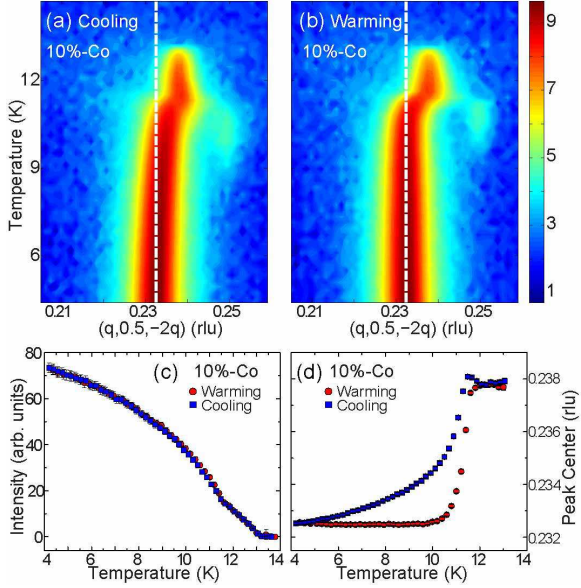}
\caption{(Color online) The $T$ dependence of the wave-vector scans of the
ICM magnetic peak for $x=0.10$ sample upon (a) cooling and (b) warming. Notice
the weak commensurate AF1 reflections at $\vec{q}=(0.25,0.5,-0.5)$ is present
near 11~K for both cooling and warming protocols. (c) and (d) display the
$T$ dependence of the integrated intensity and the peak center of the ICM
scattering.
}
\label{fig:ICM.10pCo}
\end{figure}

Figures.~\ref{fig:ICM.10pCo}(a)-\ref{fig:ICM.10pCo}(d) show the neutron
diffraction results on the $x=0.10$ sample. The hysteresis of the magnetic
order upon thermal cycling is still discernible. However, the shift of the magnetic
wave vector is much smoother upon cooling and does not show the plateau as
seen at $x=0.075$.  Accordingly, electric polarizations show
continuous growth upon cooling. The saturated $P_a$ and $P_c$ reach 100 and
30~$\rm \mu C/m^2$ at 4~K, respectively. $P_a$ exceed the maximum $b$-axis
polarization of MnWO$_4$ by nearly a factor of two. This result is consistent
with the predicted $P$ in Eqn.~(\ref{eqn:a_P}) assuming unchanged
coefficient $C_1$ in the AF2 and AF5 phases. On warming, the low-$T$
spin structure remains till 11~K and a transition to the high-$T$ collinear
AF3 phase occurs. The lack of an abrupt change in the ICM wave vector suggests the
modification of the spin structure is not as drastic as the $x=0.075$ sample
and the same ICM spin structure is preserved to the lowest temperature. In a recent
neutron diffraction study on the $x=0.10$ sample, Urcelay-Olabarria {\it et
al.} obtained a similar result and found that the magnetic spins remain in the
$ac$ plane at all temperatures while the eccentricity reduces from
$\epsilon=0.66$ at 9~K to 0.42 at 2~K, suggesting the spiral ellipse becomes
more circular at lower temperature.\cite{olabarria12} This decrease of the
eccentricity upon cooling further increases the electric polarization, as
expressed in Eqns.~(\ref{eqn:a_P}) and (\ref{eqn:c_P}). Another noteworthy
feature in the $x=0.10$ sample is that the energy of the $ac$ spiral structure
is close to other competing state including the collinear AF1 phase. This can
be appreciated by the weak magnetic reflection at the CM wave vector
$\vec{q}=(0.25,0.5,-0.5)$ near 11~K [see Figs.~\ref{fig:ICM.10pCo}(a) and
\ref{fig:ICM.10pCo}(b)].

\begin{figure}
\includegraphics[width=3.0in]{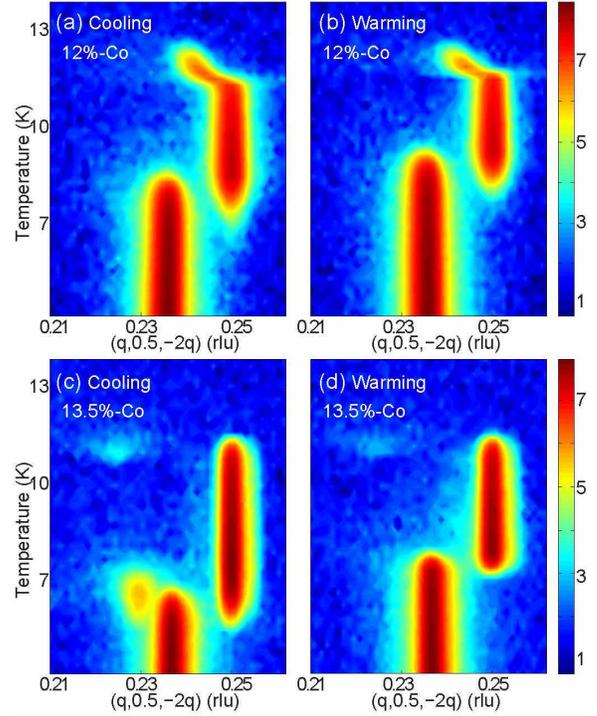}
\caption{(Color online)
The $T$ dependence of the wave-vector scans of the magnetic peaks for $x=0.12$
sample upon (a) cooling and (b) warming. The system undergoes ICM AF3, CM AF1,
and ICM AF5 phases as cooling. Similar scans on the $x=0.135$ sample are
presented on (c) and (d). Only CM AF1 and ICM AF5 orders appear as the major
phases. In (c), a minor AF2 phase with $\vec{q}=(0.23,0.5,-0.46)$
appears at 5.8~K $< T <$ 7.6~K.
}
\label{fig:ICM.12p_13p5Co}
\end{figure}

The evolution to other competing magnetic orders is manifested at
higher Co concentration. Figure~\ref{fig:ICM.12p_13p5Co} displays the
$T$ dependence of the wave-vector scans for two samples with $x=0.12$ and
$x=0.135$. At $x=0.12$, the first magnetic order present in the scattering
plane consisting of the $[1,0,-2]$ and the $[0,1,0]$ directions is the
collinear AF3 phase, which locks into the commensurate AF1 phase at lower
temperature.  With the sample cooled below 9.0~K, the previously observed AF5
phase sets in and extends to the lowest temperature without any noticeable
shift of the magnetic wave vector.  There is a narrow temperature window
in which both the AF1 and AF5 phases are present between 6.8~K $< T <$ 9.0~K.  On
warming, the reverse order of the magnetic phases is observed except the
coexisting region shifts up to 7.5~K $< T <$ 9.3~K.  Interestingly, the order
of appearance for the commensurate AF1 and the multiferroic AF5 phases is
exactly opposite to the pure MnWO$_4$, where the AF1 phase occurs at lower
temperature. At $x=0.135$, the transition between the collinear AF3 and the
AF1 phases vanishes. There is very weak scattering near the ICM wave vector
$\vec{q}=(0.225,0.5,-0.45)$ that exists in a very narrow range of
10.7~K$<T<$11.3~K, which was later identified as the AF2 phase.  The strongest
intensity is only one percent of that for the AF1 phase and is too weak to
induce any detectable polarization signal. We also observed the coexistence of
the multiferroic AF5 phase and collinear AF1 phase. The coexisting region
downshifts to 5.5~K$<T<7.2$~K for cooling and 7.0~K$<T<$7.8~K for warming. At
lower temperature, the AF1 phase is completely suppressed.

\begin{figure}
\includegraphics[width=3.0in]{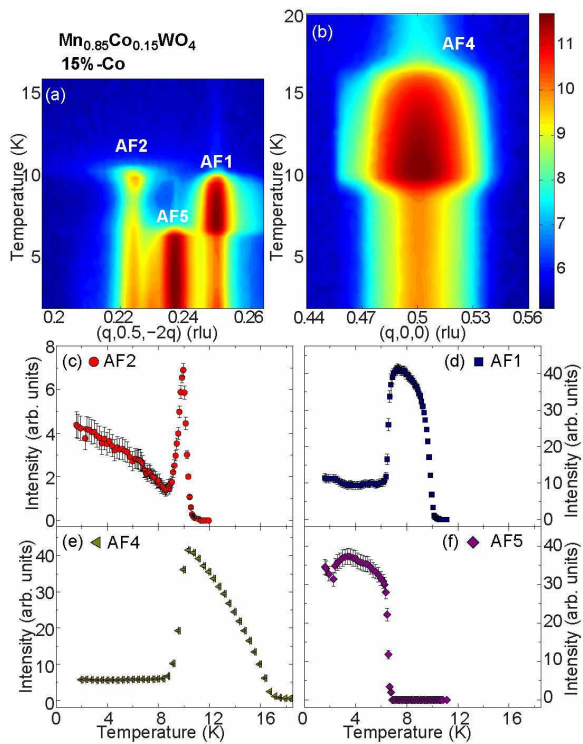}
\caption{(Color online)
The $T$ dependence of the wave-vector scan near the magnetic orders of the
$x=0.15$ sample along (a) the $[H,0.5,-2H]$ and (b) the $[H,0,0]$ directions.
(c)-(f) are the $T$ dependence of the integrated intensities for the AF2,
AF1, AF4, and AF5 phases.
}
\label{fig:ICM.15pCo}
\end{figure}

The $x=0.15$ sample is probably the most complex system with a minimum of five
coexisting magnetic phases.\cite{chaudhury10}  Figure~\ref{fig:ICM.15pCo}
summarizes the thermal evolution of the various magnetic orders probed at two
scattering planes.  Figure~\ref{fig:ICM.15pCo}(a) shows the $T$ dependence of wave-vector scans
along the $[H,0.5,-2H]$ direction where the AF1, AF2, and AF5 can be surveyed.
Figure~\ref{fig:ICM.15pCo}(b) displays the scans along the $[H,0,0]$ direction where the
commensurate AF4 magnetic order with $\vec{q}=(0.5,0,0)$ can be examined. Upon
cooling, the AF4 phase first appears around 17~K and the intensity increases
continuously till 10~K. A sharp drop in its intensity is accompanied by the
simultaneous development of the commensurate AF1 and multiferroic AF2 phases.
With the sample cooled below 6.6~K, the CM AF1 phase is also suppressed and
an additional ICM AF5 phase develops at lower temperature.
Figures.~\ref{fig:ICM.15pCo}(c)-\ref{fig:ICM.15pCo}(f) summarize the $T$ dependence of integrated
intensities for the four major magnetic (the commensurate AF1, AF4 and
incommensurate  AF2, AF5) phases in the $x=0.15$ sample. Since the magnetic
intensity of the AF2 phase closely follows the $b$ axis polarization $P_b$, it is
speculated that this state has a spiral spin structure similar to the $x \leq
0.05$ samples. Notice even at the lowest temperature, there are finite
magnetic scattering from the commensurate AF1 and AF4 phases.

\begin{figure}
\includegraphics[width=3.3in]{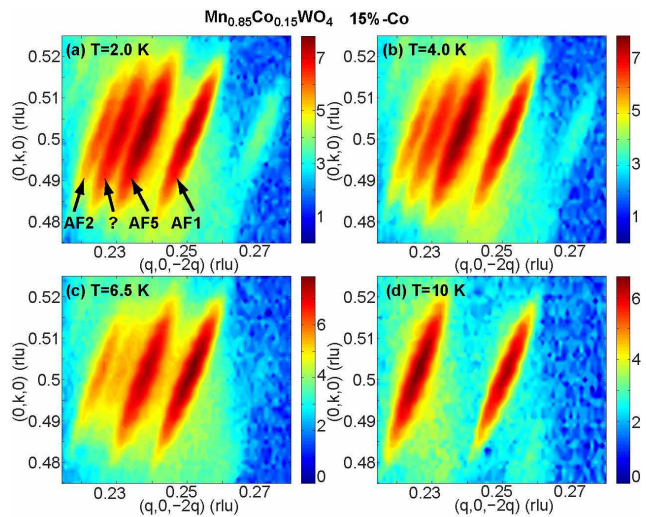}
\caption{(Color online)
The reciprocal space mapping near the CM and ICM magnetic reflections at (a)
2, (b) 4, (c) 6.5, and (d) 10~K. At least four distinct magnetic
reflections are observed in this scattering plane at low temperatures.
}
\label{fig:ICM.15pCo.map}
\end{figure}

To confirm the coexistence of various magnetic orders, we have performed
an extensive survey in the reciprocal space at selected temperatures, 2.0,
4.0, 6.5, and 10~K, as shown in Fig.~\ref{fig:ICM.15pCo.map}.  The appearance
and disappearance of competing magnetic orders is evident. At 10~K, only the
AF2 with $\vec{q}_2=(0.225,0.5,-0.45)$ and AF1 with $\vec{q}_1=(0.25,0.5,-0.5)$
phases are present, while more magnetic Bragg peaks appear at lower
temperature. A distinct magnetic reflection appearing as
the shoulder of the AF5 Bragg peak becomes visible in the low-$T$ mapping
[see Figs.~\ref{fig:ICM.15pCo.map}(a) and \ref{fig:ICM.15pCo.map}(b)]. The peak is located between the AF2 and
AF5 phases with wave vector $\vec{q}=(0.23,0.5,-0.46)$. The exact nature of
this magnetic order remains unknown since its reflection is too close to the
neighboring AF2 and AF5 Bragg peaks. This unknown magnetic phase is already
present at $x=0.135$, but only exists within a narrow temperature range
[see Fig.~\ref{fig:ICM.12p_13p5Co}(c)]. This magnetic fluctuation continues to
grow in intensity upon cooling at $x=0.15$.  Together with the already
identified commensurate AF1,AF4 phases and incommensurate AF2,AF5 phases,
there are five magnetic orders at the lowest temperature.  Such remarkable
coexistence of many competing magnetic phases marks the $x=0.15$ sample as the
most frustrated system.

Table~II lists the refined magnetic structures for the
$x=0.075,0.10,0.12$, and 0.135 samples in the AF5 phase as well as the
commensurate AF1 and AF4 phases for the $x=0.135$ sample. As mentioned in
Sec.~\ref{sec:middleX}, the $x=0.075$ sample is located near the phase
boundary between the low-$x$ spiral and the $ac$ spiral structures; its
magnetic structure can be refined as an $ac$ spiral configuration only for
$T>8$~K. In contrast, the magnetic ground states of the $x=0.10,0.12$ samples
are well described by the same $ac$ spiral structure at low temperatures.
Although the ICM magnetic order seems to be the only low-$T$ phase for the
$x=0.135$ sample, it cannot be refined by a pure $ac$ spiral structure
implying the deviation from that configuration. To get good agreement
between the observation and model calculation, a combined low-$x$
spiral and $ac$ spiral structure is chosen to fit the experimental data and yields
satisfactory result. The spin structure is best characterized as a modified $ac$
helical structure that the normal vector of the spiral plane tilts way from
the $b$ axis. Such spin order results in a reduction of electric polarization
and is consistent with the bulk measurement that both $P_a$ and $P_c$ decrease
for $x>0.10$.\cite{liang12}

\begin{table}[ht!]
\label{tab.refine.interX}
\caption{Refined parameters of the magnetic structures at $x=0.075, 0.10,
0.12$, and $0.135$. The spin configurations of the $x=0.075$ sample at 9~K and
the $x=0.10,0.12$ samples at 5~K can be refined as an $ac$ spiral structure with
its principle axes aligning along the $a$ and $c$ directions. The
$x=0.135$ sample is refined to be the AF4 phase at 12~K, the AF1 phase at 9~K
and the modified AF5 phase at 5~K. The moment direction has a angle of
$-33^{\circ}$ with respect to the $a$ axis in the AF1 phase and $-52^{\circ}$
to the $a$ axis in the AF4 phase. At the AF5 phase, the spin helix has one of
its two principle axes in the $ac$ plane and the other in the $bc$ plane.
}
\begin{ruledtabular}
\begin{tabular}{llccccc}
 $x$ & phase & \multicolumn{2}{c}{moment ($\mu_B$)}  &   & $R_{F^2} (\%)$ &	\\
    \hline
  0.135 &  AF4 	& $m_a:$ 0.817(8) &$m_c:$ -1.051(9)	&	    	& 11.33 &   \\
  0.135 &  AF1 	& $m_a:$ 3.13(3) &$m_c:$ -2.07(4)	&	      	& 8.08 &   \\ \hline
       &        & Real               & Imaginary	& $\epsilon$ 	&      &   \\
  0.075&  AF5  	& $m_a:$ 3.45(4) & $m_c:$ 2.67(4)	& 0.63(3)  	& 4.81 &   \\
  0.10 &  AF5  	& $m_a:$ 4.05(4) & $m_c:$ 3.14(5)	& 0.63(3)  	& 7.41 &   \\
  0.12 &  AF5 	& $m_a:$ 4.04(6) & $m_c:$ 3.56(7)	& 0.46(6)  	& 5.83 &   \\
  0.135 & AF5 	& $m_a:$ 4.03(4) & $m_c:$ 3.53(5)    	&	      	& 7.06 &   \\
        &      	& $m_c:$ -0.81(7) & $m_b:$ -0.10(7)  	&	      	&      &   \\
  \end{tabular}
\end{ruledtabular}
\end{table}

\begin{figure}
\includegraphics[width=3.3in]{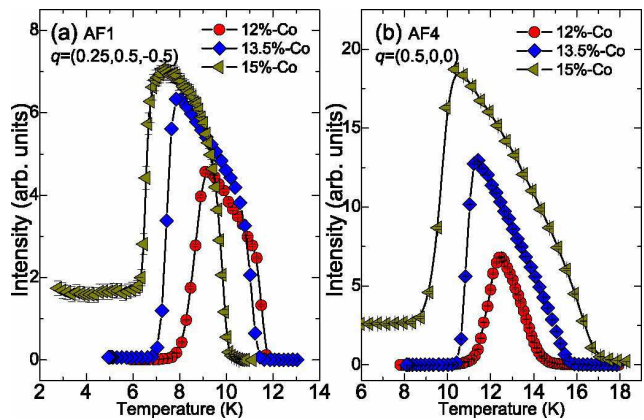}
\caption{(Color online)
The $T$ dependence of the integrated intensity for (a) the AF1 phase with
$\vec{q}=(0.25,0.5,-0.5)$ and (b) the AF4 phase with $\vec{q}=(0.5,0,0)$ at
$x=0.12,0.135$, and $0.15$. Both the AF1 and AF4 states survive at low
temperature for the $x=0.15$ sample.
}
\label{fig:OP_10p15p}
\end{figure}

With Co concentration $x>0.10$, we have observed the expansion of both
commensurate AF1 and AF4 phases as shown in Fig.~\ref{fig:OP_10p15p}.
Overall, samples in this doping region form the commensurate AF4 spin
structure with $\vec{q}_4=(0.5,0,0)$ at higher temperature and enter directly
the collinear AF1 phase upon cooling, which is different from pure MnWO$_4$.
With increasing $x$, the transition to the AF1 state moves to lower
temperature while the transition to the AF4 phase shifts to higher one.
Although both phases at $x=0.10,0.12,0.135$ are completely suppressed at
low-$T$, they survive for the $x=0.15$ sample indicating the collinear spin
structures gradually become the stable magnetic ground state at large $x$.
Magnetic structure refinements in this doping range reveal that the spin
moments in the AF4 phase are confined in the $ac$ plane, with an angle of
$-50^{\circ}$ towards the $a$ axis (see Table~II).  This spin reorientation is
again due to the strong anisotropy of Co$^{2+}$
ions\cite{forsyth94,hollmann10} that locks the Mn$^{2+}$ spins in the same
direction and makes the collinear spin structure more favorable with
increasing $x$. The spins in the collinear AF1 phase have a angle of
$-33^{\circ}$ with respect to the $a$ axis, which is different from pure
MnWO$_4$, and is probably due to the pinning of the high-$T$ AF4 magnetic
structure.

\section{magnetic order at high concentration ($0.17\leq x\leq 0.30$) }

\begin{figure}[ht!]
\includegraphics[width=3.3in]{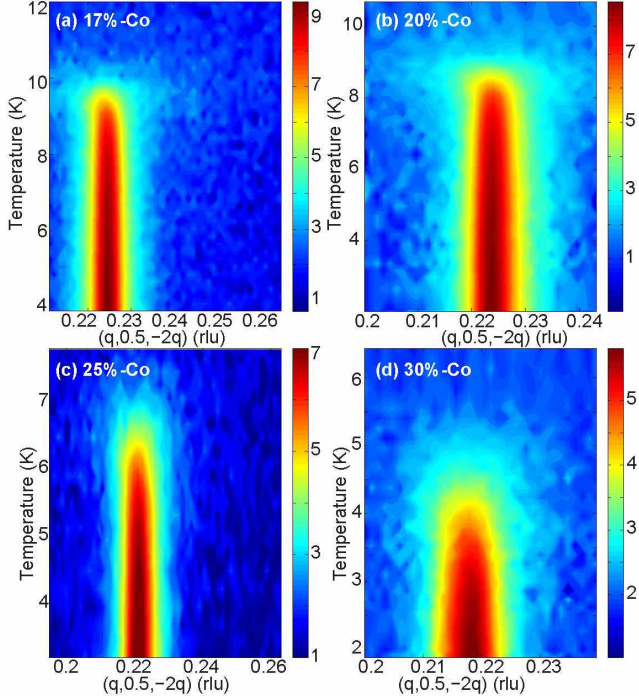} \caption{(Color
online) (a) The $T$ dependence of the wave-vector scans of the magnetic orders
along the $(H,0,-2H)$ direction for $x=0.17,0.20,0.25$, and $0.30$.
}
\label{fig:17p_30pCo}
\end{figure}

Finally, we focus on the magnetic structures for $x\geq0.17$. The bulk
polarization measurements show no detectable $P_a$ and $P_c$. Instead, the
polarization is pointing to the $b$ axis, the same direction as observed at
lower Co concentration. Thus, the switch of the polarization suggests another
major modification of the spin structure.  Figure~\ref{fig:17p_30pCo} compares
the $T$ dependence of the wave-vector scans across the ICM peak of four Co doped
samples with $x=0.17,0.20,0.25$, and $0.30$. The scattering profile in this
doping regime exhibits different character. Unlike the coexistence of various
competing magnetic orders in the intermediate doping regime, there is only one
ICM magnetic reflection at the wave vector of $\vec{q}\approx(0.22,0.5,-0.44)$.
The transition temperature decreases from $9.6$~K at $x=0.17$ to $4.4$~K at
$x=0.30$. The scattering intensity of the ICM magnetic order is also 
suppressed with increasing $x$ as displayed in Fig.~\ref{fig:OP.highX}(a).
Further survey in reciprocal space reveals one strong collinear
AF4 phase that is established at higher temperatures and persists to the lowest
temperatures [see Fig.~\ref{fig:OP.highX}(b)]. The transition temperature
increases with Co concentration and reaches 25~K for the $x=0.30$ sample. For
all samples studied, the magnetic intensities of the AF4 phase exhibit a kink
at temperatures corresponding to the onset of the mentioned ICM magnetic order
indicating the competition between the CM and ICM phases.

\begin{figure}[ht!]
\includegraphics[width=3.3in]{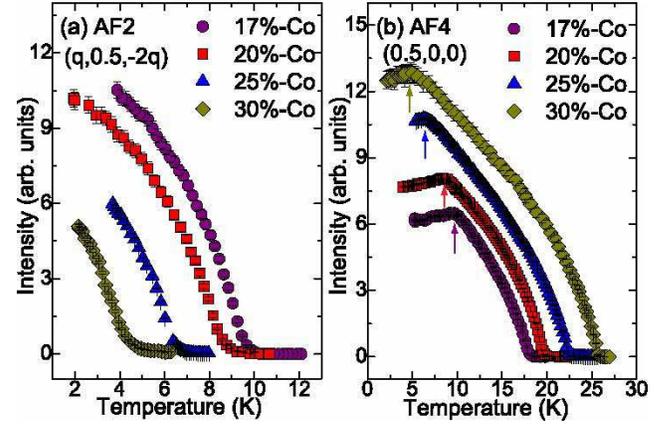}
\caption{(Color online) (a) The $T$ dependence of the integrated intensities of
the AF2 phase for $x=0.17,0.20,0.25$, and $0.30$. (b) The $T$ dependence of
the integrated intensities of the AF4 phase with $\vec{q}=(0.5,0,0)$ for the
same samples. Arrows label the transitions where the low-$T$ AF2 phases set
in. The magnetic scattering intensities are normalized to the intensities from
the nuclear reflections for comparison.
}
\label{fig:OP.highX}
\end{figure}

\begin{figure}[ht!]
\includegraphics[width=3.3in]{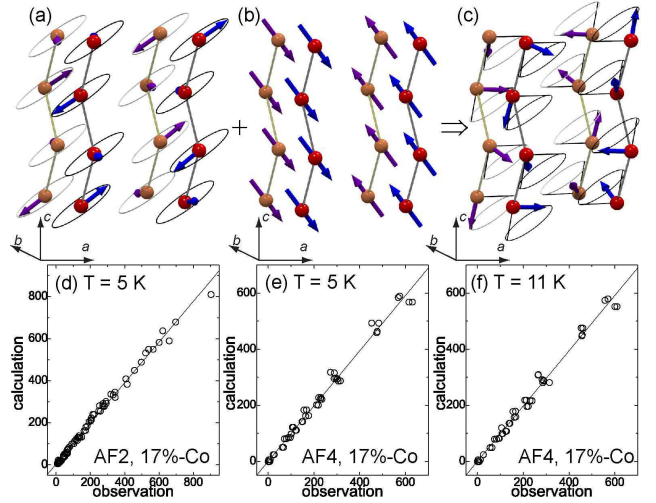}
\caption{(Color online) The spin configurations of (a) the low-$T$
incommensurate AF2 phase, (b) the high-$T$ incommensurate AF4 phase, and
(c) the conical spin order resulting from the superposition of the AF2 and AF4
phases. Agreement plots for (d) the incommensurate AF2 phase at 5~K, (e) the
commensurate AF4 phase at 5~K, and (f) the commensurate AF4 phase at 11~K.
}
\label{fig:refinement.17pCo}
\end{figure}

\begin{table}[ht!]
\caption{Magnetic structure parameters at $x=0.17$. The sample has collinear
AF4 phase with $\vec{q}=(0.5,0,0)$ at 11~K and 5~K. At $T=5$~K, additional AF2
spiral structure is formed. $m_a$ and $m_c$ are the spin components along the
$a$ and $c$ axes in the collinear phase. The real $m_{\perp b}$ and imaginary
$m_b$ denote the projected moments of the long and short spiral ellipse lying
in the $ac$ plane and along the $b$ axis. $\theta$ is the angle of the
$m_{\perp b}$ with respect to the $a$ axis.
}
\begin{ruledtabular}
\begin{tabular}{llccccc}
  phase & $T$ & \multicolumn{2}{c}{moment ($\mu_B$)}  & $\theta$  & $R_{F^2} (\%) $ &	\\
    \hline
AF4 &  11~K 	& $m_a:$ 1.47(3) & $m_c:$ -1.86(2)	& $-51.6^{\circ}$& 6.47 &   \\
AF4 & 5~K 	& $m_a:$ 1.47(1) & $m_c:$ -1.90(1)	& $-52.3^{\circ}$& 6.25 &   \\
    \hline
    &    & Real               & Imaginary	& $\epsilon$ 	&      &   \\
AF2 & 5~K 	& $m_{\perp b}:$ 2.92(7) & $m_b:$ 2.36(4)   & 0.39 	& 5.11  \\
  \end{tabular}
\end{ruledtabular}
\label{tab:refine.l7p}
\end{table}

Since all samples in this doping region show similar magnetic properties
except the transition temperature, we chose the $x=0.17$ sample for the
crystal and magnetic structure refinement and expect the other doped samples
have smooth evolution of the spin structure.  As demonstrated in
Figs.~\ref{fig:17p_30pCo} and \ref{fig:OP.highX}, there are two major magnetic
phases with ICM and CM wave vectors for $x=0.17$. We collected 193 nuclear
reflections at 5~K for the structural determination. One set of magnetic
reflections for the low-$T$ ICM magnetic structure were collected to refine
the spin structure and two sets of magnetic reflections with
$\vec{q}=(0.5,0,0)$ were collected at 5 and 11~K separately to investigate
how the collinear AF4 phase is affected by the low-$T$ ICM magnetic order.
Figures.~\ref{fig:refinement.17pCo}(a) and \ref{fig:refinement.17pCo}(b) show the corresponding magnetic
structures of the ICM spiral order and the CM AF4 phase. At 11~K, only the
collinear AF4 phase exists, the magnetic spins form in a configuration
identical to CoWO$_4$, where the moments lie in the $ac$-plane, with an angle
of $\theta\approx -50^\circ$ towards the $a$ axis.  With the sample cooled
below 10~K where the ICM order sets in, the magnetic structure of the
collinear spin order is not modified; the spins remain in the same direction
and the total moment of 2.42(4)~$\mu_B$/site at 5~K is almost the same as
2.39(3)~$\mu_B$/site at 11~K (see Table~III). On the other hand, attempting to
refine the low-$T$ ICM order using the $ac$ spiral structure gives a poor fit to
the collected data. Consequently, we adopted the spin configuration at low Co concentration
because such spin structure provides an electric polarization along the $b$
axis. As shown in Fig.~\ref{fig:refinement.17pCo}(d), this model provides a
good description of the experimental data. The spin moment of the spiral state
varies from 2.36~$\mu_B$ to 2.92~$\mu_B$, which is comparable with the moment
size at the AF4 phase. The observation of spiral order with the helix plane
similar to the low-$x$ case implies that the $x=0.17$ sample is located at
another phase boundary where the system undergoes a second spin flop
transition consistent with $\vec{P}\parallel b$. However, there is one
apparent difference between the $x\leq0.05$ and $x\geq0.17$ samples. There is
only one ICM AF2 phase in the low Co concentration samples, while the ICM AF2
phase appearing in the high-$x$ samples coexists with the commensurate AF4
phase that is established at higher temperature. The simultaneous presence of
two magnetic phases is similar to the colossal magnetoresistance related
manganite $\rm Pr_{0.7}Ca_{0.3}MnO_3$ (PCMO), in which both the ferromagnetic
and antiferromagnetic components are observed at low
temperature.\cite{yoshizawa95} This can be interpreted either as a canted
antiferromagnetic structure or coexisting FM and AFM phases. Similarly,
neutron diffraction data alone can not differentiate whether the observed
coexistence of AF2 and AF4 orders in $\rm Mn_{0.83}C_{0.17}WO_4$ arises from
two separated phases, each with distinct magnetic wave vector; or if they
originate from one single phase with two-$k$ magnetic structure.  Definitive
identification would require a spatially sensitive probe in conjunction with the
electric polarization measurement.  Our neutron diffraction result at $x=0.17$
is consistent with a recent study on the $x=0.20$ sample, where the coexisting
collinear AF4 and multiferroic AF2 phases are revealed.\cite{olabarria12b}
Those authors concluded that the superposition of the competing AF4 and AF2
magnetic structures leads to a conical antiferromagnetic order that is
depicted in Fig.~\ref{fig:refinement.17pCo}(c).  

\section{discussion and conclusions}

\begin{figure}[ht!]
\includegraphics[width=3.4in]{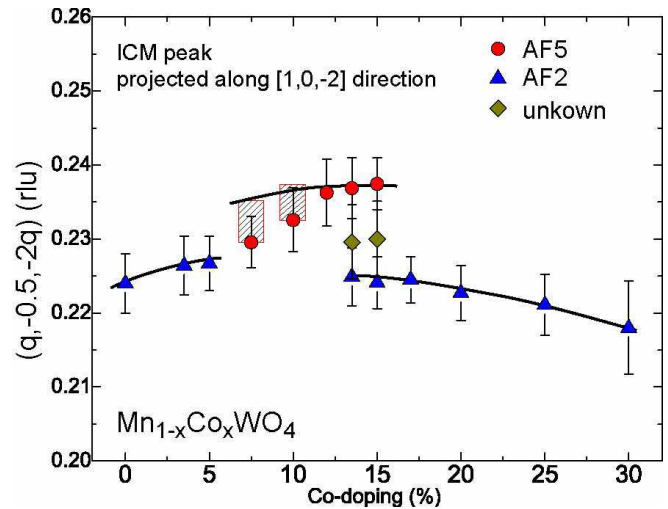}
\caption{(Color online) The doping dependence of the incommensurability of the
low-$T$ magnetic structures. The low-$T$ spin orders are labeled with
different symbols.  The shadow region for the $x=0.075$ and $x=0.10$ samples
indicates the samples undergo continuous change of the magnetic wave vectors
upon cooling. Solid lines are guides to the eye. For $x=0.15$ and 0.17, there
are additional magnetic Bragg reflection between the known AF2 and AF5 phases.
}
\label{fig:ICM}
\end{figure}

The comprehensive single-crystal neutron diffraction measurement, in
combination with the magnetic property and polarization measurements, make it
possible to construct the phase diagram of the $\rm Mn_{1-x}Co_{x}WO_4$ as
a function of Co concentration and temperature. Unlike other transition
metal ion doped MnWO$_4$, where only one type of spin configuration is
stabilized, a rich variety of spin structures and complex evolution between
different phases are observed as the Co concentration is increased.
The spin anisotropy of the Co ions  plays a vital role in
defining the low-$T$ magnetic structures. Although confined in the same $ac$
plane, the spin easy axis in CoWO$_4$ is $-45^{\circ}$ from the $a$ axis, and
is nearly 90$^{\circ}$ away from the easy axis direction in
MnWO$_4$.\cite{weitzel70,forsyth94} With increasing $x$, the long axis of the
spiral ellipse that initially has a positive angle towards the $a$ axis will
tilt gradually to the negative direction due to the single ion anisotropy of
the Co$^{2+}$.  The rotation of the spin helix plane leads to a decrease of the
electric polarization $|\vec{P}|$ that is compatible with the magnetic
symmetry, as well as the energy gain that is coupled to the $|\vec{P}|$. At
the critical concentration of $x=0.075$, the system cannot gain enough energy
to maintain the multiferroic phase, thus results in the spin flop transition.
The new $ac$ spiral structure helps the system lower the free energy because of
the large value of the ferroelectric polarization.  For $0.075\leq x
\leq0.15$, the presence of the $P_a$ and $P_c$ instead of $P_b$ is consistent
with an $ac$ spiral spin structure, and highlights the significant intrachain
as well as the interchain interactions.  Although the samples in the
intermediate doping exhibit similar spiral structure, the maxima of the
electric polarizations only occur near $x=0.10$.  It results mainly from the
rotation of the spiral plane as exemplified by the neutron diffraction data
from the $x=0.075, 0.135$ samples. The perfect $ac$ spiral order is realized at
$x\approx0.10$ that induces the largest polarization. Further increasing $x$
not only causes the deviation from a pure $ac$ spiral configuration, but also
introduces collinear AF1 and AF4 spin orders that reduce the effective moment
of the spiral structure responsible for the polarization. A second spin-flop
transition takes place with $x \geq 0.17$ and leads to a similar spiral
structure as in the low Co concentration. This phase coexists with a CM
collinear AF4 phase with spin configuration similar to CoWO$_4$.  The gradual
suppression of the electric polarization in this doping region is mainly
caused by the increasing collinear AF4 order, other than the rotation of the
spiral plane, as observed at $x \leq 0.05$.

The phase diagram is characterized by three well-defined regions distinguished
by different spin spirals.  However, the evolution of the magnetic spin
structure within an individual region is gradual. This can be better
appreciated by examining the concentration dependence of the
incommensurability of the low-$T$ noncollinear order, as shown in
Fig.~\ref{fig:ICM}. The magnetic wave vector of the ICM order does not exhibit
a lock-in value over the wide range of Co concentration. Instead, it varies
smoothly within each of the three regions.  For the samples near the $x=0.075$
phase boundary, the ICM magnetic structure is unstable such that a small
temperature variation will cause the rotation as well as the periodicity
change of the spiral structure. All these observations reinforce that the
magnetic structure results from the delicate balance between the competing
exchange interactions and spin anisotropy of the transition metal ions.

Without chemical substitution, MnWO$_4$ appears to be a highly frustrated
system where the magnetic and ferroelectric properties can be modified by
external stimuli like the magnetic and electric
field.\cite{sagayama08,taniguchi08a,taniguchi08b,taniguchi09,finger10a,finger10b,nojiri11}
The introduction of Co ions with distinct spin anisotropy provides another way
to fine tune the magnetic ground states.  This is similar to the
rare-earth multiferroic manganite $R$MnO$_3$, in which the ferroelectric
polarization is enhanced by the magnetic order of rare earth
elements,\cite{prokhnensko07} except that the tuning parameter is on the same
magnetic site in the case of $\rm Mn_{1-x}Co_{x}WO_4$. We hope the current
experimental study will inspire further theoretical effort to understand the
magnetic and ferroelectric order parameters in this doped system.  Most
importantly, such work will provide a new pathway to design and synthesize
magnetoelectric-control materials with multiple magnetic and ferroelectric
ground states.

We thank R. S. Fishman, C. de la Cruz and B. Chakoumakos for invaluable
discussions.  The work at ORNL is supported by the Division of Scientific User
Facilities of the Office of Basic Energy Sciences, US Department of Energy.
Work at Houston is supported in part by the T.L.L.  Temple Foundation, the
John J. and Rebecca Moores Endowment, and the State of Texas through TCSUH,
the US Air Force Office of Scientific Research, Award No. FA9550-09-1-0656,
and at LBNL through the US DOE, Contract No. DE-AC03-76SF00098.

%\bibliography{MnWO4}
%

\end{document}